# ASYMP: Fault-tolerant Mining of Massive Graphs

EDUARDO FLEURY, Google
SILVIO LATTANZI, Google
VAHAB MIRROKNI, Google
BRYAN PEROZZI, Google

We present ASYMP, a distributed graph processing system developed for the timely analysis of graphs with *trillions* of edges. ASYMP has several distinguishing features including a robust fault tolerance mechanism, a lockless architecture which scales seamlessly to thousands of machines, and efficient data access patterns to reduce per-machine overhead. ASYMP is used to analyze the largest graphs at Google, and the graphs we consider in our empirical evaluation here are, to the best of our knowledge, the largest considered in the literature.

Our experimental results show that compared to previous graph processing frameworks at Google, ASYMP can scale to larger graphs, operate on more crowded clusters, and complete real-world graph mining analytic tasks faster. First, we evaluate the speed of ASYMP, where we show that across a diverse selection of graphs, it runs Connected Component 3-50x faster than state of the art implementations in MapReduce and Pregel. Then we demonstrate the scalability and parallelism of this framework: first by showing that the running time increases linearly by increasing the size of the graphs (without changing the number of machines), and then by showing the gains in running time while increasing the number of machines. Finally, we demonstrate the fault-tolerance properties for the framework, showing that inducing 50% of our machines to fail increases the running time by only 41%.

## 1 INTRODUCTION

As a fundamental tool in modeling and analyzing social and information networks, processing large-scale graphs is an important component of any tool set for big data analysis. In order to handle huge graphs with *trillions* of edges, such a tool needs to distribute the graph and its computation across many machines. Building such a distributed graph processing tool as a general-purpose library is feasible only if the distributed framework has the following basic properties: (i) it supports *fault-tolerace* against failure or preemption of a subset of machines, (ii) it runs on a shared cluster with commodity hardware, and (iii) it provides a simple and general way for its users to implement new algorithms on this framework. Without these properties, a distributed programming framework is not going to be successful or widely used. For example, without a robust fault-tolerant mechanism, running any distributed algorithm requires both a dedicated set of high-performance machines, and the careful monitoring of its execution. This is not desirable for a general-purpose tool that is meant for running distributed data mining applications programs. In fact, the success and widespread use of MapReduce [6], Hadoop, and Pregel [18] can be attributed to satisfying these properties.

Most of the existing big data analysis and graph mining distributed frameworks with these properties (e.g., MapReduce and Pregel) are based on the concept of performing synchronized "rounds" of computation in parallel. While such synchronous computation is suitable for a range of applications, it incurs certain inefficiency in CPU usage and may result in longer running time due to the synchronization step. These inefficiencies are especially noticeable in graph mining tasks where the computation may be dependent on topological properties of the graph (e.g, runtime dependent on its diameter [5, 13]), and therefore require many rounds of expensive iteration.

In this paper we introduce ASYMP, a new distributed processing framework for asynchronous message-passing graph computation which addresses these issues. Following previous distributed graph processing frameworks, ASYMP provides a simple abstract language for message-passing-based algorithms among nodes of an underlying graph. In addition, ASYMP has several distinguishing features that makes it scalable and efficient for certain applications: For example, ASYMP only keeps a smaller state space for the nodes of the graph in memory, and this makes ASYMP scalable for graphs with trillions of edges; moreover, ASYMP implements a robust and effective fault-tolerant mechanism that makes it suitable for running on thousands of machines on the shared cluster with commodity hardware; finally, asynchronous computation makes ASYMP's performance significantly better than other graph computation frameworks. In order to show ASYMP's superior performance, we benchmark graph mining primitives such as computing connected components and single-source shortest path and report the scalability of our algorithms under various resource constraints and failure patterns. We also compare it to other distributed processing frameworks such as MapReduce [20], MapReduce plus distributed hash-table [14], and Pregel [18].

Our main contributions in this paper are to introduce a new distributed graph processing framework and its fault-tolerant implementation, and present an extensive empirical study for this framework in terms of its scalability, efficiency, and fault-tolerance compared to other frameworks. Specifically, we analyze ASYMP's:

**Speed**: We show that a simple implementation of the connected components in ASYMP outperforms the best algorithms known for connected components in MapReduce [14, 20], and Pregel [18]. Following company regulations, and similar to previous research [14, 18], we do not report absolute running times, but we do report relative improvements over the best previously published work.

**Scalability**: We illustrate ASYMP's scalability in both computational resources and in relation to the size of its input graph. In the first experiment, we keep the input constant while varying the number of machines available. We observe that up to using a certain number of machines, the running time decreases proportional to the number of machines. In the second experiment, we keep the resources constant while using a series of increasingly larger hyper-link graphs as input. To the best of our knowledge, the web graphs we consider are the largest graphs ever analyzed in the literature. We show that the running time scale linearly with the size of the input — even for graphs which have over a trillion edges.

**Fault Tolerance**: We examine the fault-tolerance of ASYMP by artificially making groups of machines fail over the course of running the algorithm. We observe that ASYMP can handle an excessive number of failures by recovering the states, e.g, making 50% of machines fail while running the code increases the running time by only 41%.

## 2 RELATED WORK

Large-scale graph mining has important commercial applications, and so a number of computational models have been proposed for the distributed setting. In this section we discuss how ASYMP compares to previously proposed models from the literature.

**MapReduce** MapReduce is a general paradigm for distributed data processing [6]. It implements an excellent fault-tolerance scheme, and can be used for graph processing if it is run iteratively in several synchronized rounds. In practice, this required synchronization imposes a high cost, and can render algorithms which require many rounds of computation infeasible. For example, a straightforward MapReduce algorithm to find connected components [5, 13, 19] grows a Breadth First Search (BFS) tree from each node in parallel. This approach requires $O(d)$ iterations in the worst case, where $d$ is the diameter of the largest connected component, and therefore does not scale well to even moderate diameter graphs.

Unlike MapReduce, ASYMP is a specialized graph processing tool that can be used to implement iterative message passing algorithms in a very efficient way. Our carefully controlled experiments show that it can significantly outperform state of the art MapReduce algorithms. The ASYMP implementation of connected components runs up to *50 times faster* than the most efficient connected components algorithm in MapReduce [20].
**Bulk Synchronous Parallel (BSP):** The BSP paradigm [23] is used by recent distributed graph processing systems like Pregel [18] and Giraph [4]. Pregel uses the concept of vertex processors which can send messages to one another. Computation proceeds in a series of synchronized 'super-steps', where all active vertices (those with pending messages) are executed at once, and then vertices exchange messages. While BSP is generally considered more efficient for graph processing, it has been shown [20] that MapReduce algorithms can have better latency in congested clusters (owing to their superior fault tolerance.)

Unlike Pregel, ASYMP has no global synchronization step. Instead, the operations of sending and receiving messages are split into two methods that are executed asynchronously and in arbitrary order. This difference results in not only more effective CPU usage by ASYMP, but also affords an improved fault tolerance scheme (as an entire super step's computation is not blocked by a single machine failure).
**Asynchronous Message Passing Frameworks:** More recently, there has been a general trend towards introducing asynchronicity in message passing frameworks. GraphLab [17] and PowerGraph [10] have been proposed for distributed graph processing using asynchronous message passing to further improve running times of algorithms. Furthermore, several more recent asynchronous graph processing models have been proposed [24, 25]. Finally we note that recently many more frameworks have introduced asynchronous computation to improve graph processing, for example PrIter [26], Maiter [27], and GiraphUC [11] (See [22] for a survey).

This branch of the literature is the most directly related to our work, and we note that ASYMP was developed in parallel with much of the early research in this area (see [9] for a prior public disclosure of ASYMP).

However, ASYMP has several differentiating features. First ASYMP *does not store* the edges of the graph in memory—instead it fetches them from disk whenever needed. This optimization greatly reduces the memory overhead associated with processing web-scale graphs, which is especially important when running in clusters whose resources may be overcommitted. Second, in its checkpointing system, ASYMP carefully ensures that only the last message of a failed node to and from its neighbors is replayed. This implementation lets ASYMP recover from failures faster and more effectively compared to other frameworks. Finally, and most importantly, many previous asynchronous frameworks (e.g. GraphLab [17] or PowerGraph [10]) still rely on a set of locks to ensure all operations are serializable. However, ASYMP does not offer that guarantee. Not depending on these locks allows the message delivery to be optimized and machine failures to be handled more efficiently. This ultimately results in the ability to scale to larger graphs, operate on more crowded clusters, and complete real-world graph mining analytic tasks faster.

## 3 THE ASYMP MODEL

The goal of ASYMP is to solve problems that can be expressed on a graph and can be solved by exchanging messages over its edges. In this section, we begin by describing ASYMP's abstract model and the details of its worker architecture. Next, we remark on the nature of supported algorithms for ASYMP. Finally, we discuss ASYMP's fault tolerance and message queuing systems.

### 3.1 Abstract Model

The framework has two phases that happen in a sequential way: "propagation" and "merger". In the propagation phase, all the vertices of the user-provided graph are distributed among many machines and then user-defined

```
 1: for v ∈ G do                                              // Loop 1. Initialization
 2:     activate_v ← Initialization(v)
 3:     if activate_v then
 4:         S ← S ∪ {v}                                        // Add to active set
 5:     end if
 6: end for
 7:
 8: while S ≠ ∅ do                                            // Loop 2. Process active vertices
 9:     v ← S.next()
10:     N_v ← FetchNeighbors(v)                               // Retrieve adjacency list of v
11:     for w ∈ N_v do
12:         m_w = CreateMessage(w)
13:         Q.add(m_w)                                         // Add to message queue
14:     end for
15:     S ← S\{v}                                             // Deactivate node v
16: end while
17:
18: while Q ≠ ∅ do                                            // Loop 3. Process pending messages
19:     m_w ← Q.next()
20:     DeliverMessage(m_w)
21:     activate_w ← RecieveMessage(m_w)
22:     if activate_w then
23:         S ← S ∪ {w}                                        // Add to active set
24:     end if
25: end while
```

Fig. 1. Outline of the ASYMP Propagation Phase

messages begin to be sent between different vertices until all vertices agree that there are no more messages to be sent. In other words, the whole framework stops when the graph converges to a stable state. When the graph has converged to a stable state the merger phase starts. This stage allows user-defined code to read the final state of each vertex to extract the application output.

An ASYMP program consists of four main components: an *initialization step*, a component for *creating messages* for active nodes, a component for *delivering messages on a priority queue*, and finally *executing and possibly activating* on the target nodes. These programs are executed in parallel across many machines by the ASYMP framework.

Figure 1 shows a sketch of three loops which roughly describe the framework's primary logic. The first loop (lines 1-6) calls the initialization component of the user program, which defines the initial state of each node, and activates a subset of nodes to start the computation. After initialization, the next two loops are executed in parallel until the stopping condition of the user program is met (i.e. there are no active nodes remaining). The second loop (lines 8-16) is responsible for dealing with active nodes and creating messages for them based on an order dictated by a priority queue. The third loop (lines 18-25) is responsible for delivering the messages, executing the target nodes by updating their state, and possibly marking the target nodes as active.

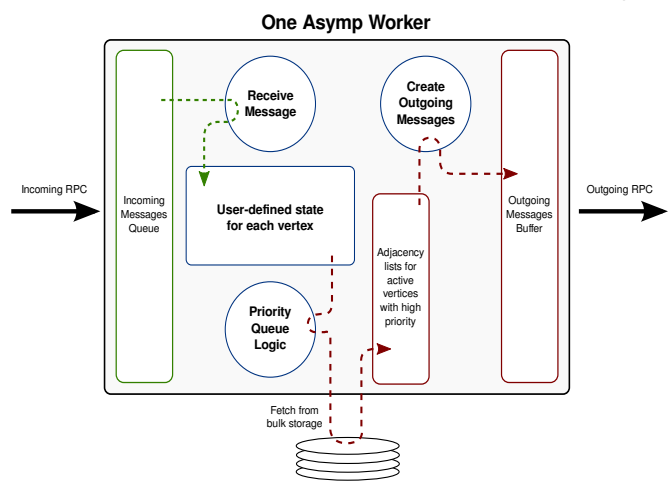

Fig. 2. Architecture of one ASYMP worker.

## 3.2 Five-Thread Model

After initialization, each ASYMP worker uses five parallel threads to implement the asynchronous messaging described in the previous section (Figure 1, lines 8-25). The communication of these five threads occurs through a series of message queues. The architecture of an ASYMP worker running these five threads is depicted in Figure 2.

**Active node selector.** The first thread is in charge of seeking the next active node to execute its message-creation step. This thread enforces a specific priority queue rule for picking the next active node. It then writes to a list of adjacency lookups which is in charge of fetching adjacent neighbors of active nodes.

**Fetching Adjacency lists.** The next thread is in charge of fetching the adjacency list of nodes and storing them in memory for next action. This thread is very I/O intensive since it sends a large amount of data over the network.

**Creating outgoing messages.** The third thread is in charge of creating the messages to be sent to neighbors. This thread reads the adjacency lists of nodes created in memory by the previous thread and executes the user specified code to produce outgoing messages for the neighbors. This is the main piece of code that the user of the ASYMP framework needs to write in order to develop a new algorithm in the framework. After creating all messages from a node $v$ this thread also de-actives node $v$. Finally this thread writes all messages created in this step into a queue kept in memory.This thread is CPU intensive as it runs the main piece of the code that is in charge of creating the message sent around nodes.

**Sending messages to neighbors.** The fourth thread is in charge of delivering the messages created in the previous thread to the neighbors. Messages are not sent to another machine one by one. This thread groups into batches organized by target machine. Then it creates an RPC call to a different server transmitting a group of these messages to another machine. This thread is I/O intensive as it sends data over the network to be picked up by another worker.

**Processing incoming messages.** The fifth thread is in charge of executing the user code on the target vertex. It reads messages from the inbox queue and for each message it executes the code specified by the user of the ASYMP framework. The user code may also decide to activate the node that receives the message. Finally this thread updates the state of the nodes that receives a message and may add them to the priority queue of active nodes. This thread is CPU intensive as it runs the user specified code for receiving the message and changing the state of the nodes accordingly.

### 3.3 Supported Algorithms

ASYMP's asynchronous nature requires that implemented algorithms support processing messages which may arrive in an arbitrary order. Failure to take this into account when designing an algorithm for ASYMP can result in an implementation which does not converge, or that yields non-deterministic results. For example, consider the well-studied PageRank algorithm. In it, each node's PageRank is defined as the sum of its neighbors scores. Since multiple messages from a single neighbor may arrive consecutively, a correct implementation of PageRank in ASYMP requires additional state (storing the latest contribution from each of a node $v$'s neighbors – instead of just the sum).

Conversely, algorithms where messages are idempotent and supersede previous messages are guaranteed to run correctly in ASYMP. One way to achieve this property is to make sure that an algorithm satisfies the self-stabilizing property [8]. There is a large body of work on such self-stabilizing algorithms, which covers a diverse set of problems, including connected components, shortest paths, label propagation and many more [8, 12]. In this paper we focus on connected components and shortest paths for simplicity.

Finally, we remark that while developers should be aware of the effects of asynchronous messaging, it has actually been shown to provide improved performance on a number of tasks (such as faster convergence when solving linear systems [1] and optimizing neural networks [21]).

### 3.4 Fault-Tolerance

Production clusters are often shared among many users running various tasks. Task failure is frequent in these crowded environments due to both preemption from higher priority jobs, and the failure of unreliable hardware. Having strong support for fault-tolerance is, therefore, a critical part of any distributed graph mining framework. ASYMP is designed with this scenario in mind, and implements a high level of fault-tolerance by *asynchronously* storing the state of nodes on disk so that we can recover to a more recent state in case of failures or preemptions. More specifically, it implements fault tolerance in three steps:

**Writing checkpoints** In preparation for eventual failures, each ASYMP worker periodically (and asynchronously) saves the state of its vertices from memory to disk. This work is done independently within each worker and allows the computation to be resumed from an earlier state in case of a failure.

**Recovering itself** Whenever a worker fails it is rescheduled (possibly on a different machine). Once that worker is functional again its first step is to look for the checkpoint files it may have written in the previous steps. It then reads such files and restores the state of its vertices.

**Requesting lost messages** At this stage while the worker has recovered its own state, it still misses any messages that it may have received since that last checkpoint was written. To overcome this the failed worker sends a message to all other machines specifying at what time it began losing messages. Each peer then starts a recovery process where it re-sends to the failed machine all the messages that had been created by its vertices since the failure. Once every peer has completed the recovery process the failed machine is guaranteed to be up to date and can continue to exchange messages until the computation is over.

The above process does not require interference from the ASYMP user. The framework handles checkpointing and replaying of messages automatically.

### 3.5 Message Order Optimization

In order to optimize the performance of the ASYMP system further, we can deploy different priority rules for the CreateMessage priority queue. The goal of the priority queue is to reduce the total number of messages sent in the system, by executing active vertices in a specific order rather than arbitrarily. The best order to execute vertices will be algorithm-dependent, and so ASYMP allow implemented algorithms to define their vertex priority as a function of each vertex's state. We examine the effect of message order optimization on an implementation of the connected components algorithm in Section 5.6.

## 4 ASYMP PROGRAMMING ENVIRONMENT

This section elaborates on aspects of ASYMP's C++ API.

**Data Types.** ASYMP supports a number of different data types to maintain the state for each vertex, including POD(Plain Old Data), Protocol Buffers and ordinary strings.

**Initialization.** Upon starting, ASYMP executes the method Init() once for each vertex of the graph. The user is expected to set the initial state of each vertex and possibly activate it by using two methods: MutableValue() and RequestPropagation().

**Message Passing.** Due to the asynchronous nature of ASYMP, users have to implement two different methods to send and receive messages. The first method is CreateOutgoingMessages() which is executed on each active vertex. The user is expected to read the current vertex state, parse the provided adjacency list, and then call SendOutgoingMessages() as needed.

Conversely, as messages arrive at the target machine ASYMP executes the second method ReceiveMessage(), which allows the user to inspect the incoming message, mutate the state and possibly activate the vertex receiving it. CreateOutgoingMessages() will eventually be executed on this vertex, as explained above. Messages will be exchanged until the system converges to a state where no vertices are active and there are no messages in the system.

**Output extraction.** As vertices are no longer active and the propagation cycle is deemed complete the framework initiates a second phase. In it, the fourth method GetOutputString() is executed for each vertex. This allows the user to inspect the final state of each vertex and generate an output string from it.

**Input and Output.** Two inputs must be provided: the Initialization Table and the Adjacency List. Both must be SSTables where each key is an unique string identifying one vertex of the graph. The values in the first table and second tables will be forwarded to user code whenever Init() or CreateOutgoingMessages() are executed. Values can be arbitrary strings as they will be parsed by user code.

Finally the output of an ASYMP computation is an SSTable where the keys are the ids of each vertex and the values are those created in GetOutputString().

### 4.1 Examples

In this subsection we give the pseudo-code implementation of two fundamental algorithms in the ASYMP framework: connected components and single-source shortest path.

**Connected Components.** Recall that given a graph $G(V, E)$, a connected component $S$ is a maximal subset of nodes of the graph for which there exists a path from each node $u \in S$ to any other node $v \in S$. The idea of the algorithm is the following: Each node initiates its state as its ID, and the goal is for each node to compute the minimum ID of a node in its connected component. One can simply do so by propagating the new state of each node to its neighbors, and upon receiving a message from neighbors, a node updates its state to the new state if the new state is less than its current state. If the state is updated, the node sends its new state to its neighbors. This process continues until no node changes its state at which time each node already computed the ID of the minimum ID node in its connected component. The ASYMP pseudo-code is shown in Figure 3.

**Single-source Shortest Path.** Here, we discuss the algorithm for the single-source shortest path problem. In this problem, given an edge-weighted graph $G(V, E)$ and a source $s$, this problem is to compute for each node node $u$, the length of the shortest path from $s$ to $u$. If all the edge weights are positive, the Dijkstra algorithm for the shortest path problem can be easily implemented in ASYMP: the state of each node is the current computed distance from node $s$. The states are initialized at $\infty$ for all nodes except $s$ which has the initial state of 0, and node $s$ sends its state to its neighbors. The update rule is also simple: node $u$, upon receiving a new state from a neighbor

**Vertex State:** `int64 cluster_id;`
**Initialization:**
```
cluster_id = node_id;
RequestPropagation();
```
**CreateMessage:**
```
foreach neighbor u:
    SendMessage(u, cluster_id);
```
**ReceiveMessage:**
```
if (received_id < cluster_id):
    cluster_id = received_id;
    RequestPropagation();
```

Fig. 3. Connected Components

**Vertex State:** `double dist_source;`
**Initialization:**
```
if (node_id == source_id) {
    dist_source = 0;
    RequestPropagation();
} else dist_source = M;
```
**CreateMessage:**
```
foreach neighbor u:
    SendMessage(u, dist(node_id, u)+dist_source);
```
**ReceiveMessage:**
```
if (received_distance < dist_source):
    dist_source = received_distance;
    RequestPropagation();
```

Fig. 4. Shortest Path

$v$, examines if its current state is larger than $w(v, u) + \text{state}(v)$, and if so, $u$ updates its state to $w(v, u) + \text{state}(v)$ and sends this state to its neighbors. The program is shown in Figure 4.

## 5 EMPIRICAL EVALUATION

In this section, we present an empirical evaluation of our ASYMP implementations. We ran all of our jobs multiple times on a shared production cluster with commodity hardware, and report the median running time in the results below. We first describe our experimental design and then report our empirical results.

Specifically, we show that ASYMP is:

(1) **Fast** - simple algorithms implemented in ASYMP perform *5-50 times* faster when compared against optimized implementations in MapReduce and Pregel.

(2) **Scalable** - ASYMP's runtime scales linearly with the size of the input graph, and it can take advantage of increasing computational resources.

(3) **Fault Tolerance** - inducing 50% of ASYMP's workers to fail only increases the running time by 41%.

### 5.1 Experimental Design

**Datasets**: We ran our algorithms on three categories of graphs, as shown in Table 1.

| Network | # nodes | # edges |
|---|---|---|
| Patent citation [16] | 3,774,768 | 16,518,948 |
| Twitter Followers [15] | 35,689,148 | 41,652,230 |
| LiveJournal authorship [16] | 4,847,571 | 68,993,773 |
| Friendster friendship [16] | 65,608,366 | 1,806,067,135 |
| UK Webgraph [2] | 105,896,555 | 6,603,753,128 |
| Google+ | 177,878,516 | 2,917,041,952 |
| Related entity | 31,386,984 | 3,773,675,522 |
| Orkut social | 157,546,418 | 17,474,363,130 |
| Document similarity | 4,697,876,565 | 452,638,340,365 |
| WebGraph1 | ∼ 16B | ∼60B |
| WebGraph3 | ∼47B | ∼180B |
| WebGraph9 | ∼110B | ∼540B |
| WebGraph27 | ∼356B | ∼1,620B |
| RMAT 26 | 32,803,311 | 2,103,850,648 |
| RMAT 28 | 121,228,778 | 8,472,338,793 |
| RMAT 30 | 447,774,395 | 34,044,283,063 |
| RMAT 32 | 1,652,692,682 | 136,596,059,559 |

Table 1. Sizes of the graphs used for algorithm evaluation.

- **Public Graphs:** A collection of publicly available graphs including a LiveJournal authorship graph [16], a Twitter user follower graph [15], a Friendster friendship graph [16], a patent co-authorship graph [16], and a snapshot of the UK web subgraph [2].
- **Google Graphs:** To illustrate real-world performance, we also evaluated our framework on many internal networks associated with various Google projects including Orkut and Google+[1], document and entity similarity graphs, and a sequence of increasingly larger web graphs, the largest of which has 356B nodes and 1.62T edges.
- **Random Graphs:** Finally, we also evaluate on a collection of RMAT benchmark graphs that are randomly generated graphs with increasing number of nodes and edges[3]. Specifically, we generate graphs with a maximum of $2^{26}, 2^{28}, \ldots, 2^{32}$ nodes and expected degree of 32 using RMAT parameters $(a, b, c, d)$ =(0.47, 0.19, 0.19, 0.05). RMAT is a recursive model of randomly generating graph with several desirable properties such as power-law degree distribution property, small world property, and inclusion of many dense bipartite subgraphs. To generate an RMAT graph, one recursively subdivides the adjacency matrix in four equal quadrants and elects to recurse on one of the four quadrants with unequal probability $(a, b, c$ or $d)$.

The graphs we analyze are very diverse, with varying degree distribution, diameter and network structure. To illustrate this, we show the distribution of connected component sizes for WebGraph1, UK Web, Patents, Orkut, and RMAT30 in Figure 5. It is obvious that the distributions are quite different - for example, the patent graph does not have a giant connected component and most vertices are part of mid-size components.

**Dataset Diversity:** To underline the different complexity between different datasets we consider, Table 2 shows the number of messages sent by the ASYMP framework to compute connected components for different networks.

**Execution Environment**: All experiments were conducted in a live production environment, and are indicative of real world performance.

**Baselines**: Limitations on running third-party code on our distributed environment permit only a formal comparison of ASYMP with our internal tools, MapReduce and Pregel.

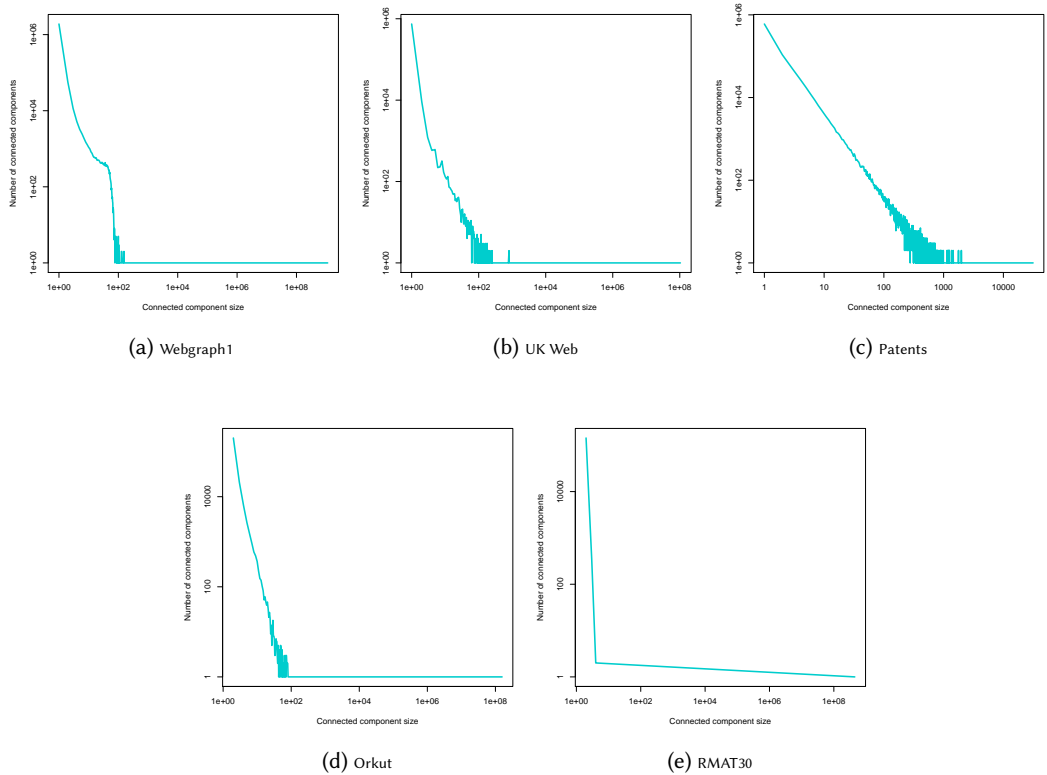

Fig. 5. Distribution of connected component sizes for select graphs. The graphs we consider vary greatly in their topology.

**Evaluation Metrics**: Consistent with company policy, we report only relative (and not absolute) statistics. For example, when evaluating run time, we will report relative speed-up compared to the slowest running approach.

### 5.2 Speed Evaluation

In order to evaluate the speed of ASYMP, we compare a straightforward implementation of connected components (section 4.1) against the following state of the art baselines:

- **MR-external**: The best publicly available baseline for connected components (i.e., the Hash-to-Min algorithm [20]).
- **MR-internal**: Our best *internal* implementation of connected components in MapReduce.
- **MR+DHT**: Our best implementation of connected components in MapReduce with distributed hash-table service or DHT [14][2].
- **Pregel**: Our best implementations of connected components [7] in Pregel [18][3].

---

[2]We remark again that the MapReduce or MapReduce+DHT implementations in [14] apply several algorithmic techniques, and load balancing tricks – however our implementation in ASYMP is the simplest possible implementation of connected components.

[3]For our Pregel implementation, we have several variants of differing complexity. Here, we used the algorithm in [7], as it had the best running time.

| Network | # messages |
|---|---|
| Patent citation [16] | 30,067,956 |
| Twitter Followers [15] | 1,710,961,215 |
| LiveJournal authorship , [16] | 157,801,362 |
| Friendster friendship [16] | 3,688,012,811 |
| UK web subraph [2] | 7,301,990,682 |
| Google+ | 3,568,260,903 |
| Related entity | 5,971,414,353 |
| Orkut social | 17,655,951,864 |
| Document similarity | 553,108,466,266 |
| WebGraph1 | 150,143,527,632 |
| WebGraph3 | 647,548,736,826 |
| WebGraph9 | 1,184,961,687,811 |
| WebGraph27 | 5,489,624,632,323 |
| RMAT 26 | 4,566,673,292 |
| RMAT 28 | 14,611,113,008 |
| RMAT 30 | 59,770,908,427 |
| RMAT 32 | 142,686,070,643 |

Table 2. Number of messages passed for the Connected Components algorithm on each graph.

All of the algorithms were run on the same shared cluster, with the same computational resources (virtual machines, memory, etc.). The relative speed-up of running these algorithms is presented in Figure 6.

On all graphs, a simple implementation of the connected components in ASYMP runs substantially faster than all MapReduce algorithms, regardless how efficient they are. Specifically we see in Figure 6a that the ASYMP implementation is between *3 to 50 times* faster than MR-external, the best publicly available MapReduce implementation from [20].

In general, ASYMP is also much faster than the optimized Pregel implementation. The only graph where Pregel is competitive is the Twitter follower graph, which has remarkably few edges (average degree of 1.1), and both implementations converge in very few iterations. We contrast that with the performance on RMAT graphs in Figure 6b. The RMAT graphs are much more difficult cases, where the entire graph forms a single giant connected component. On these large graphs (which require prolonged iteration), we see that ASYMP is *6 to 10 times* faster than an optimized Pregel implementation.[4]

### 5.3 Input Scalability

To investigate the scalability of our algorithms while increasing the size of the graphs, we turn to the webgraphs and to the synthetically generated RMAT graphs and report the relative running time of our code compared to the smaller graphs in the same family. Our goal is to confirm that the running time of these algorithms grow linearly as a function of the size of the graphs. We also examine the increase in the number of messages sent and CPU and RAM usage as the size of the graph increases. We provide these numbers both for connected components and single-source shortest path ASYMP implementations. Before describing the results in details, we note that we expect the number of messages to be a function of the size of the graph, the topology of the network, effectiveness of the priority queue, and the number of failures. Furthermore, we expect the RAM usage to be a function of the size of graph and of the running time and finally, the CPU usage to be a function of the graph size only.

---

[4]Slower MapReduce implementations are omitted in Figure 6b to show speed-up relative to Pregel.

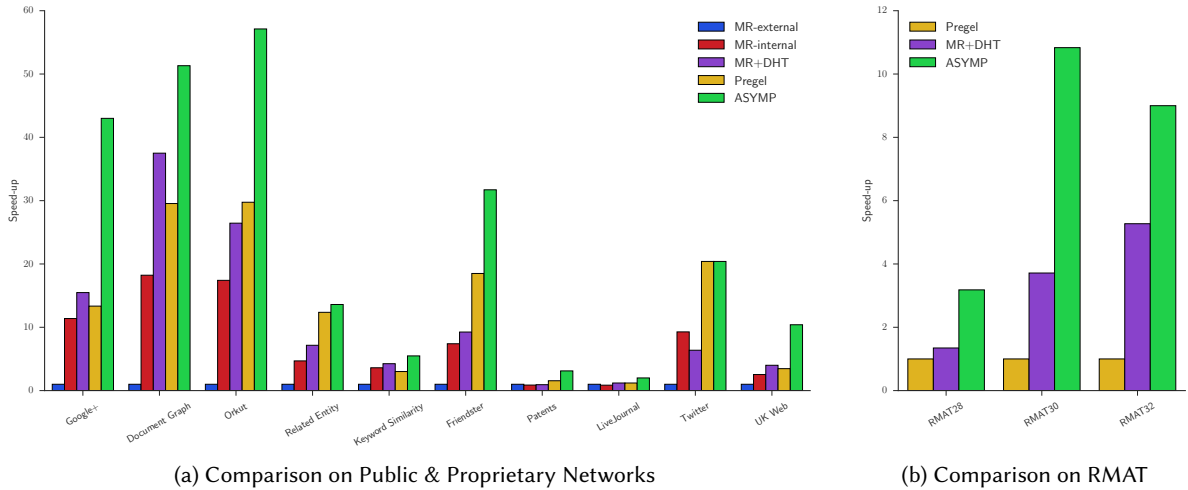

(a) Comparison on Public & Proprietary Networks

(b) Comparison on RMAT

Fig. 6. Relative speed-up of the connected component algorithm on real (a) and synthetic (b) networks.

In the first experiment, we considered RMAT graphs of increasing sizes, and run the connected components algorithm (Section 4.1) with the same number of machines. As shown in Figure 7a the running time scales sub-linearly at the beginning and then linearly in the size of the input graphs, in fact the running time increase by only a factor of 8 while the size of the graph increases by a factor of 70. It is interesting to note that, as expected, CPU usage, RAM usage and number of messages scale linearly with the size of the graph. It is particularly interesting that thanks to the priority queue the large increase in number of messages sent does not slow the running time of the algorithm.

Now we turn our attention to our second scalability experiment, in this case we run the same algorithm but on Webgraph snapshots of different sizes. Also in this case(Figure 7b) we can note that the running time increase sub-linearly initially and then linearly in the number of nodes while the other 3 measures increase roughly linearly in the number of nodes in the graph.

We also observe that the increase in the number of messages is bigger for Webgraph than for RMAT, this can be explained by the fact that solving connected components on Webgraphs is harder, since RMAT graphs tend to have lower diameter, and the task of computing connected components for them become easier.

Finally, we run a scalability test for the single-source shortest path algorithm from Section 4.1. In Figure 7c, we see that the shortest path algorithm has even better scalability properties than the connected component algorithm. In fact, we see that all our four metrics scale initially sub-linearly and only at the end linearly in the number of nodes. In particular, using the same number of machines and resources, while increasing the size of the graphs 27 times, the running time increases only by an order of 8 times.

## 5.4 Parallelizability

To investigate the dependency of ASYMP on the amount of resources available, we run our code on the same graph while increasing the number of allocated machines gradually. In particular, we study ASYMP's parallel scalability and its sensitivity to increasing resources by running connected components and single-source shortest path on Webgraph27 and RMAT32 graphs while increasing the number of allocated machines.

In Figure 8a, we summarize the experimental results for the RMAT32 graph when we increase the resources by a factor 2, 4, 8, 16, 32 and 64. Figures 8b and show the experimental results for connected components and single-source shortest paths for the Webgraph27 while increasing the available machines by a factor of 2 and 3.

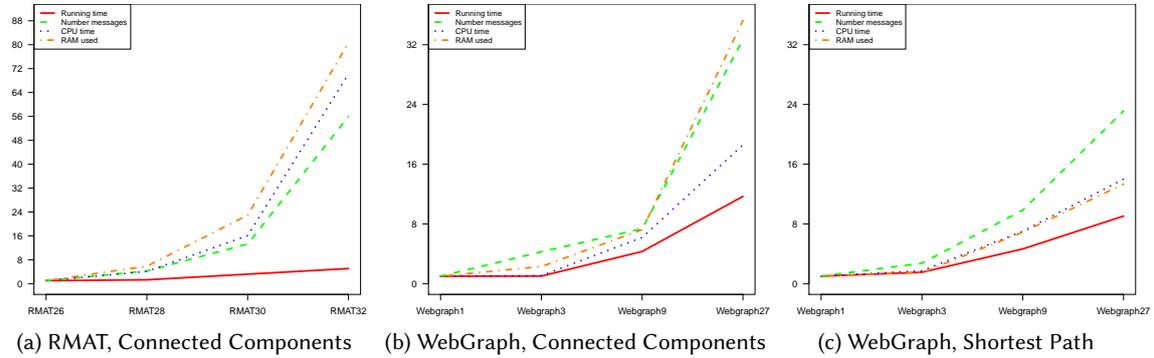

(a) RMAT, Connected Components  (b) WebGraph, Connected Components  (c) WebGraph, Shortest Path

Fig. 7. Running time comparison of algorithms for graphs of increasing size. (y-axis shows relative running times)

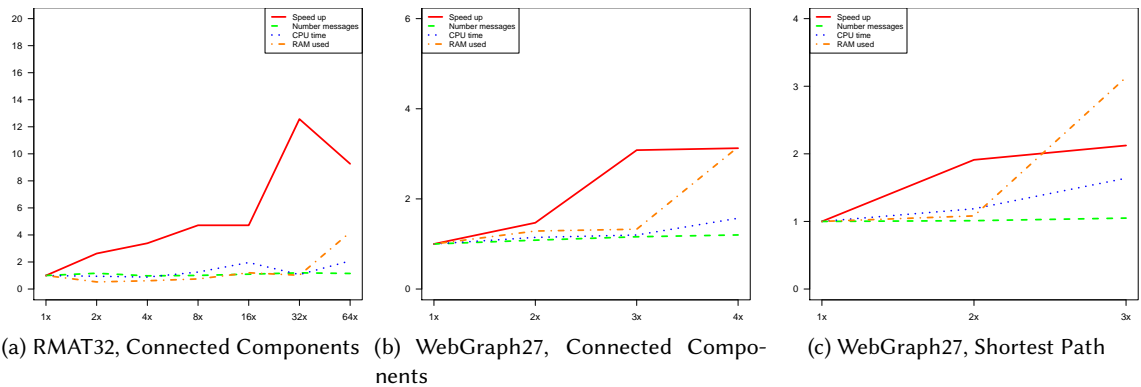

(a) RMAT32, Connected Components  (b) WebGraph27, Connected Components  (c) WebGraph27, Shortest Path

Fig. 8. Running time comparison of algorithms with increasing resources. (y-axis shows relative speed up – higher is better)

We observe similar trends in investigating all these experiments. Note that in these plots, instead of the relative running times, we report the speedup of the code for more machines, i.e., for $m$ machines, we report the ratio of the running time for $x$ machines over the running time with $mx$ machines. Therefore, the higher the number is, the better the running time.

We first elaborate on our observations for the RMAT32 graphs, but most of these observations hold for Webgraph27 as well. First of all, initially we obtain a speed-up that is almost linear with the number of machines that we use, but later we do not see as much benefit in increasing the number of resources. In fact by doubling the number of machines, we obtain a speed-up of roughly a factor less than 3, and by increasing the resources by a factor of 4 we obtain a speed-up of almost a factor of 4. Secondly, when we increase the number of machines from x32 to x64 the running time of the code increased due to some overhead incurred by increasing the number of machines. This overhead cost has the following three fundamental reasons:

(i) The priority queue operates independently within each machine. As we split the work among more machines, we start seeing local optimizations instead of global ones, i.e. messages may be deemed important by a given machine (and propagated) even though they are irrelevant to the global optimization.

(ii) There are fixed costs associated with starting and finishing the propagation, i.e., the more machines we add, the more time we need to spend to set up these machines, and the closer the running time gets to this lower boundary of setting up all machines.

(iii) The number of failures or preemptions may increase so the framework could lose a lot of time in recovering from those preemption or failures.

Thirdly, while the running time decreases by the number of machines, the number of messages sent does not increase with the number of machines. This is not surprising because, as mentioned before, the number of messages is a function of: (i) the size of the graph, (ii) the order in which we send messages and (iii) the number of failures we have. In other words, in the best case scenario the system would propagate only on the edges of the spanning forest and the total number of messages would be roughly equal to the number of edges. The actual numbers are worse than that because the prioritization is not perfect, but it is still good enough to keep the number of messages linear to the number of edges of the graph.

Finally, we note that as the number of machines increases the used RAM increases but the CPU usage does not increase as much. This happens because CPU cost is mostly a function of the size of the graph (edges and vertices), and therefore increasing the number of machines does not affect it significantly (except for the increased communication overhead and worse prioritization of messages which is done independently in each machine). On the other hand, RAM usage has an important component which is a function of number of machines times the running time. From the x32 case to the x64 case, the number of machines doubled without a corresponding decrease in running time, therefore the RAM cost increased. Note the same happened between x8 and x16.

In Figure 8b, we run connected components on Webgraph27 dataset and analyze our metrics when we increase the resources by a factor of 2, 3 and 4. Also in this case, we observe similar trends: initially there is almost a linear speedup when the number of machines increases by a factor of 2 or 3, but then when we increase the number of machines from a factor of 3 to a factor of 4, we do not observe a significant speedup due to the overhead kicking in. Interestingly also in this case we observe a spike in RAM usage when we increase the number of machines without obtaining a significant speedup.

Finally, we run also our single-source shortest path code on the Webgraph27 dataset, and analyze our metrics when we increase the resources by a factor 2 and 3. In this case we get a linear speed up only when we double the number of machines then we can already observe a sublinear improvement and a spike in RAM usage. Similar to the previous cases, even though, the running time decreases and CPU and RAM usage increase a bit, it is worth noting that the number of messages stays roughly the same during the experiment. Also RAM usage increased a lot from x2 to x3. This is due to having almost the same running time, while increasing the number of machine by 50%.

## 5.5 Fault-Tolerance

In order to study the performance of ASYMP under failure, we artificially force a subset of machines to fail with specific frequencies, and examine the increase in the running time as a function of the failure patterns. We implement this by grouping machines into batches, and performing rolling failures such that a specified number of machines fail throughout the job. Specifically, we examine cases where (50%, 100% and 200%) of the machines fail. This causes half the machines to fail once, all the machines to fail once, and all machines to fail twice, respectively. The results of this study for the connected components algorithm on the Orkut social network are presented in Figure 9a.

First, we note that thanks to its asynchronous nature, ASYMP is resilient under extreme stress conditions. In fact, even when all failures are induced on all machines at least twice during the connected components computation, the running time increases only of a factor of 4 or 5. This is particularly impressive because after each failure the machine has to recover its entire state. Another positive property of ASYMP is that the running time increases sublinearly in the number of failures in the system. For example, the blue and the green line in figure 9a are less than a factor of two away. Furthermore, when we double the percentage of failures we observe sublinear increase in the running time.

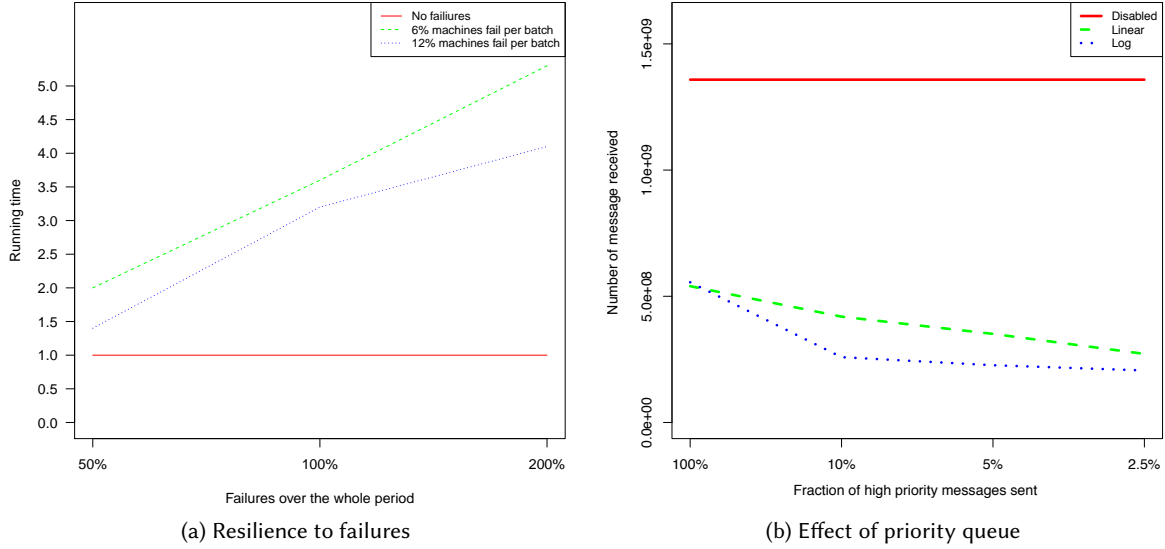

(a) Resilience to failures  (b) Effect of priority queue

Fig. 9. Analysis of different failure patterns (a) and priority queue strategies (b) when running connected components on Orkut.

### 5.6 Message Order Optimization

In the connected components algorithm we want to propagate messages from vertices that belong to clusters with very low IDs sooner than later because they have the greatest potential of moving the target vertex into a cluster where they will remain until computation is done. Here, we study the impact of enabling ASYMP's message order optimization on the number of messages transmitted during the connected components algorithm on the Orkut graph. We examine the following three priority queues:

**Disabled**: No priority queue, vertices are activated in an arbitrary order which does not depend on their current cluster ID.

**Linear**: Every node gets a priority that is proportional to their current `cluster_id`. We bucket IDs in non-uniform buckets of increasing size. This is useful because is reserves more precision for the lower end of the cluster ID range allowing for the distinction between clusters with low or very low IDs.

**Log**: We define the priority as: $P = \frac{\log(\text{first eight bytes of vertex id})}{\log(1 << 64)}$. This is a generalization of the Linear priority queue, which yields a priority between 0 and 1, where 0 means "propagate sooner".

For all priority queue strategies, we have several runs where in each run we change how strongly to enforce a given priority queue. At any given time ASYMP propagates only a fraction of high-priority messages; the lower this fraction is, the more we enforce the priority queue ordering, since if we push a large fraction of them at the same time, we are implicitly ignoring the ordering.

In Figure 9b, we show the results when ASYMP propagates 100%, 10%, 5% or 2.5% of the messages. Applying a priority queue more strongly decreases the number of messages accepted by each vertex. We also note that the continuous generalization of the logarithmic priority queue slightly outperform the linear strategy.

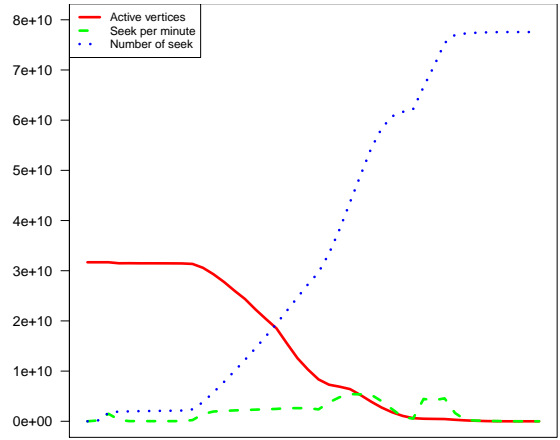

Fig. 10. The evolution of the number of messages sent by the ASYMP framework during a run of the connected component algorithm.

### 5.7 Evolution of #messages sent over time

The goal of this section is give a sense of running a message-passing algorithm in ASYMP. Figure 10 depicts the information that was aggregated among all ASYMP workers throughout the message propagation phase. Its vertical axis has been normalized with respect to the total number of vertices in the input graph.

As expected, at the beginning of the propagation phase of connected component algorithm, 100% of the vertices are active since they request propagation in their initialization. This percentage decreases over time until reaching zero thus indicating the algorithm has converged. The green line represents the rate at which ASYMP fetches the adjacency lists of the active vertices and provide them with user code so that messages can be created and sent. Different factors affect this rate. Among them we have (1) the number of active vertices, (2) the read speed of the media holding the data and finally (3) the availability of CPU to process the adjacency lists that have already been loaded into memory. Note in Figure 10 , the rate is very low in the first quarter of the propagation and then becomes higher for the second and third quarters. That increase in the seek ratio is the result from an increase in the bandwidth available for I/O operations. It is also possible to see that during the third quarter, there are two spikes in the seek ratio. These indicate the moment where several workers had to resend messages to help recover machines that had failed. Finally, the seek ratio was very low in the last quarter given that only few vertices were still active and many workers are mostly idle waiting for stragglers to complete their work.

By accumulating the values from the previous data series, we can calculate the total number of times vertices sent messages to their neighbors. During a run of the algorithm, each vertex was selected for propagation 2.5 times, on average. Note that the minumum number of propagation per node would be 1 (if we could have access to a spanning forest). The difference between the minimum number and what we have is due to two reasons: Firstly, the sequence of propagation is non-optimal, meaning vertices propagate messages with a given cluster and then have to resend messages once they realize they moved elsewhere, and secondly, machines failed requiring some work to be redone.

## 6 CONCLUSIONS

In this paper we introduced ASYMP, a new large-scale graph processing framework based on asynchronous message passing. The main distinguishing features of this framework compared to other frameworks are (i) computation is done via asynchronous message passing among nodes, (ii) only nodes (and not edges) are loaded into main memory, and edges are stored on disk and are fetched whenever needed, (iii) an effective fault-tolerant method is implemented by asynchronously storing the latest states of the nodes, and (iv) the order and frequency of sending messages can be controlled via a priority queue on each machine. We showed the effectiveness of this framework by depicting its scalability, its parallelism, its superior fault-tolerance, and finally, by showing that a simple implementation of connected components in ASYMP achieves a significant speedup compared to other graph processing frameworks including Pregel, and iterative MapReduce.

**Acknowledgments.** We would like to give special thanks to Jerry Ding, Nissan Hajaj, and Ivan Serna for their critical help in performing this research. We also thank Hossein Bateni, Raimondas Kiveris, and the rest of Google NewYork graph mining team for their input and their support.

## REFERENCES

[1] Dimitri P Bertsekas and John N Tsitsiklis. 1989. *Parallel and distributed computation: numerical methods*. Vol. 23. Prentice hall Englewood Cliffs, NJ.

[2] Paolo Boldi, Massimo Santini, and Sebastiano Vigna. 2008. A Large Time-Aware Graph. *SIGIR Forum* 42, 2 (2008), 33–38.

[3] Deepayan Chakrabarti, Yiping Zhan, and Christos Faloutsos. R-MAT: A Recursive Model for Graph Mining. In *ICDM 2004*.

[4] Avery Ching and Christian Kunz. 2010. Giraph : Large-scale graph processing on Hadoop. In *Hadoop Summit*.

[5] Jonathan Cohen. 2009. Graph Twiddling in a MapReduce World. *Computing in Science and Engineering* 11, 4 (2009), 29–41.

[6] Jeffrey Dean and Sanjay Ghemawat. 2008. MapReduce: simplified data processing on large clusters. *Commun. ACM* 51 (2008). Issue 1.

[7] Frank K. H. A. Dehne and Silvia Gtz. 1998. Practical Parallel Algorithms for Minimum Spanning Trees.. In *SRDS*.

[8] Shlomi Dolev. 2000. *Self-Stabilization*. MIT Press.

[9] E. Fluery, S. Lattanzi, and V. Mirrokni. 2014. ASYMP: Fault-tolerant Graph Mining via ASYnchronous Message Passing. In *GraphLab Conference*. http://thegraphlabconference2014.sched.com/speaker/vahab_mirrokni.1sjif08b

[10] Joseph E. Gonzalez, Yucheng Low, Haijie Gu, Danny Bickson, and Carlos Guestrin. PowerGraph: Distributed Graph-Parallel Computation on Natural Graphs. In *OSDI 2012*. https://www.usenix.org/conference/osdi12/technical-sessions/presentation/gonzalez

[11] Minyang Han and Khuzaima Daudjee. 2015. Giraph Unchained: Barrierless Asynchronous Parallel Execution in Pregel-like Graph Processing Systems. *PVLDB* (2015).

[12] Tetz C. Huang. 2005. A self-stabilizing algorithm for the shortest path problem assuming read/write atomicity. *J. Comput. Syst. Sci.* 71, 1 (2005), 70–85. https://doi.org/10.1016/j.jcss.2004.12.011

[13] U. Kang, Charalampos E. Tsourakakis, and Christos Faloutsos. 2009. PEGASUS: A Peta-Scale Graph Mining System- Implementation and Observations. (2009). http://citeseerx.ist.psu.edu/viewdoc/summary?doi=10.1.1.156.764

[14] Raimondas Kiveris, Silvio Lattanzi, Vahab Mirrokni, Vibhor Rastogi, and Sergei Vasiliviski. 2014. Connected Components in Map-Reduce and Beyond. (2014).

[15] Haewoon Kwak, Changhyun Lee, Hosung Park, and Sue Moon. 2010. What is Twitter, a Social Network or a News Media?. In *(WWW)*.

[16] Jure Leskovec and Andrej Krevl. 2014. SNAP Datasets: Stanford Large Network Dataset Collection. http://snap.stanford.edu/data. (June 2014).

[17] Yucheng Low, Danny Bickson, Joseph Gonzalez, Carlos Guestrin, Aapo Kyrola, and Joseph M. Hellerstein. 2012. Distributed GraphLab: a framework for machine learning and data mining in the cloud. *Proc. VLDB Endow. 2012* (2012). http://dl.acm.org/citation.cfm?id=2212351.2212354

[18] Grzegorz Malewicz, Matthew H. Austern, Aart J.C Bik, James C. Dehnert, Ilan Horn, Naty Leiser, and Grzegorz Czajkowski. 2010. Pregel: a system for large-scale graph processing. In *SIGMOD*.

[19] Steven J. Plimpton and Karen D. Devine. 2011. MapReduce in MPI for Large-scale Graph Algorithms. *Special issue of Parallel Computing* (2011).

[20] Vibhor Rastogi, Ashwin Machanavajjhala, Laukik Chitnis, and Anish Das Sarma. 2013. Finding Connected Components in Map-Reduce in Logarithmic Rounds. In *ICDE*.

[21] Benjamin Recht, Christopher Re, Stephen Wright, and Feng Niu. 2011. Hogwild: A lock-free approach to parallelizing stochastic gradient descent. In *Advances in Neural Information Processing Systems*. 693–701.

[22] Nadathur Satish, Narayanan Sundaram, Md. Mostofa Ali Patwary, Jiwon Seo, Jongsoo Park, M. Amber Hassaan, Shubho Sengupta, Zhaoming Yin, and Pradeep Dubey. 2014. Navigating the maze of graph analytics frameworks using massive graph datasets. In *SIGMOD 2014*. 979–990.

[23] Leslie G. Valiant. A bridging model for parallel computation. *Commun. ACM* (????). https://doi.org/10.1145/79173.79181

[24] Guozhang Wang, Wenlei Xie, Alan J. Demers, and Johannes Gehrke. 2013. Asynchronous Large-Scale Graph Processing Made Easy. In *CIDR 2013*.

[25] Chenning Xie, Rong Chen, Haibing Guan, Binyu Zang, and Haibo Chen. SYNC or ASYNC: time to fuse for distributed graph-parallel computation. In *ACM SIGPLAN 2015*.
[26] Yanfeng Zhang, Qinxin Gao, Lixin Gao, and Cuirong Wang. 2013. PrIter: A Distributed Framework for Prioritizing Iterative Computations. *IEEE Trans. Parallel Distrib. Syst.* 24, 9 (2013), 1884–1893.
[27] Yanfeng Zhang, Qixin Gao, Lixin Gao, and Cuirong Wang. 2014. Maiter: An Asynchronous Graph Processing Framework for Delta-Based Accumulative Iterative Computation. *IEEE Trans. Parallel Distrib. Syst.* 25, 8 (2014), 2091–2100.